# Electronic energy loss processes for slow H and He ions in metals and insulators: new insights

*D. Goebl, D. Roth and P. Bauer*

Institut für Experimentalphysik, Abteilung für Atom- und Oberflächenphysik, Johannes Kepler

University Linz, Altenberger Straße 69, 4040 Linz, Austria

**Abstract:**

Electronic stopping of H and He ions in metals and insulators is analyzed at velocities below 0.2 a.u., i.e. below 1 keV for H and below 4 keV for He. In metals, stopping of H ions is affected by *d*-electrons only when the *d*-band extends up to the Fermi energy; for He ions, also *d*-bands well below the Fermi energy contribute significantly to electronic stopping. In insulators, the low threshold velocity for electronic stopping cannot be explained by electron-hole pair excitation; charge exchange cycles, however, may govern the threshold behavior of electronic stopping in ionic crystals.

When ions transverse matter they lose energy along their trajectory according to the stopping power $S = dE/dx$. By convention, one distinguishes between the electronic stopping power $S_e$ (energy transfer to the target electrons), and nuclear stopping power $S_n$ (energy transfer to target nuclei). To eliminate the trivial dependence of $S_e$ on the atomic number density $n$, one can introduce the electronic stopping cross section (SCS) $\varepsilon_e = 1/n\, S_e = \int T\, d\sigma$. Here, $T$ denotes the transferred energy, $d\sigma$ is the corresponding cross section and the integral accounts for the summation over all electronic excitation processes. In addition to electron-hole pair excitation, also ionization, plasmon excitation, and electron promotion processes in atomic collisions (e.g., charge exchange cycles as observed for He ions in Al [1]) are included.

The physics involved in electronic stopping of light ions is very well understood for high velocities, $v \gg v_F$, where $v_F$ denotes the Fermi velocity [2,3]. At ion velocities $v < v_F$, however, there is still need for deeper understanding of the prevailing mechanisms. In this regime the projectiles only interact with weakly bound electrons in the valence or conduction band of the target system. When these electrons are modeled as a free electron gas (FEG), the stopping power is known to exhibit velocity proportionality [4, 5, 6, 7],

$$S_e = Q(Z_1, r_s)\, v. \qquad (1)$$

Here, the friction coefficient $Q$ is a function of the atomic number of the projectile $Z_1$, and the Wigner-Seitz radius of the FEG, $r_s = (3/4\pi n_e)^{1/3}$, with the FEG density $n_e$. Since the early days of ion physics, substantial theoretical efforts were devoted to modeling of $Q$ for different projectiles, by use of a variety of approaches to describe the response of the target electrons to the ion charge, e.g., dielectric theory [4,5,8] or density functional theory (DFT) [6,9]. In a thorough analysis [10], it was demonstrated that Eq. (1) holds for ion velocities up to $v \leq v_F$, and the experimental data are well described by a friction coefficient deduced from the DFT model [6], when the materials are characterized by an effective FEG density, obtained from experimental plasmon losses [11].

In different classes of materials interesting features in $\varepsilon_e$ have been revealed by recent energy loss experiments for H and He ions at velocities $v < 0.5$ atomic units (a.u. $m_e = \hbar = e = 1$; all quantities are given in atomic units, unless otherwise noted). Deviations from velocity proportionality were observed for metals with a distinct excitation threshold, e.g. for materials with *d*-bands located several eV below the Fermi energy $E_F$ [12,13,14,15,16,17]. These observations were traced back to the fact that significant excitation of *d*-electrons is only possible if the projectile velocity is sufficiently high, i.e. when it exceeds a distinct velocity threshold. As a consequence, the stopping power is proportional to the ion velocity only if $v$ is smaller than a certain "kink velocity" $v_k$ or sufficiently large so that all valence or conduction electrons contribute to the stopping process [10]. A thorough theoretical study of this phenomenon was conducted by Zeb et al. [18,19]. These investigations revealed that even at $v < 0.2$ a.u. a noticeable contribution to electronic stopping of H and He projectiles originates from excitation of electrons below the *d*-band offset.

In insulators, the band gap constitutes an excitation threshold, below which electronic stopping due to electron-hole pair excitation vanishes. Although their band gap energies (KCl, SiO$_2$: 8 eV, LiF: 14 eV) are much larger than the *d*-band offsets in noble metals, experimental values for the threshold velocities $v_{th}$ in insulators are found to be smaller than the kink velocities $v_k$ in noble metals [20,21,22]. A strong perturbation of the band structure due to the presence of the ion is expected [23], but a clear explanation for the low value of $v_{th}$ has not been given so far [19,20]. To find an interpretation, these observations may be related

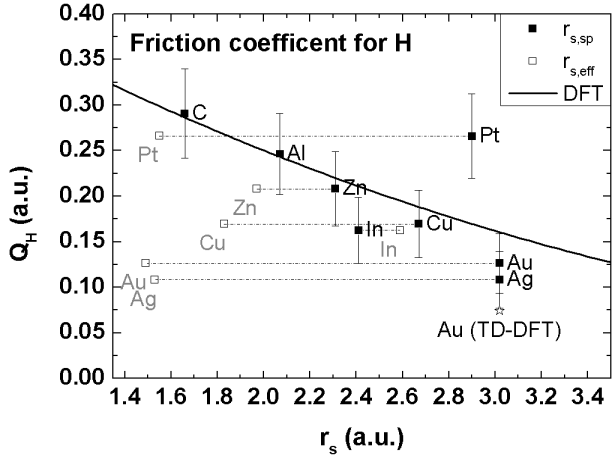
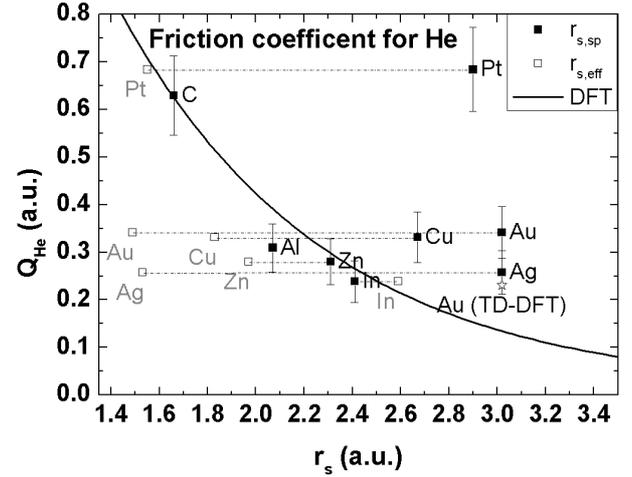

**FIG 1** *Friction coefficient for H ions, $Q_H$, as a function of $r_s$. Full symbols correspond to $r_{s,sp}$ values based on sp-electrons only, open symbols represent effective $r_{s,eff}$ values based on experimental plasmon energies [11]. Carbon data was taken from [24]. Solid line indicates the friction coefficient predicted by DFT calculations [26,27,28]. The asterisk corresponds to data obtained in TD-DFT calculations [18]*

**FIG 2** *Friction coefficient for He ions, $Q_{He}$, as a function of $r_s$. Carbon data was taken from [29,30]. Full symbols refer to $r_{s,sp}$ values based on sp-electrons only, the open symbols represent effective $r_{s,eff}$ values deduced from experimental plasmon energies [11]. Solid line indicates the friction coefficient predicted by DFT calculations [26,27,28]. The red asterisk corresponds to data obtained in TD-DFT calculations [18]*

to the electronic properties of large band gap insulators, like electronic defects (e.g. excitons) or ionicity of the chemical bond and the Madelung potential for ionic crystals.

In the first part of this investigation we deal with electronic stopping of slow light ions in metals, the second part addresses the threshold behavior of electronic losses in ionic crystals.

To analyze electronic stopping in metals at low velocities, we discuss the friction coefficients $Q_H$ and $Q_{He}$ at velocities below $v_k$, i.e. below 0.2 a.u., as a function of $r_s$. Fig. 1 shows $Q_H$ for different materials, deduced from recent experiments in Ag, Al, Au, Cu, In, Pt and Zn [1,16,17,25]. Full symbols refer to $r_{s,sp}$ values corresponding to the density of sp-electrons in the conduction band, open symbols are based on the effective values $r_{s,eff}$, which adequately describe $\varepsilon_e$ at $v \approx v_F$.

The solid line corresponds to the predictions of the DFT model [26,27,28]. It describes the $r_s$-dependence of the experimental data very well in a wide $r_s$-range (1.6 to 3.0 a.u.), when the $r_{s,sp}$-values are employed. The use of $r_{s,eff}$ leads to excellent agreement only for Pt, for which the d-electrons represent a significant part of the electron density at $E_F$. This finding can be interpreted in the following way: unless the d-electrons contribute to the density of states at $E_F$, electronic stopping can be adequately described by using the sp-density only. Fig. 1 includes also the friction coefficient deduced from a recent study of electronic stopping of channeled ions in gold using time-dependent (TD-) DFT calculations (asterisk) [18]. There, it was shown that already at low ion velocities electrons significantly below $E_F$ can be excited efficiently. These findings can be reconciled with our results by assuming that these excitations originate from the sp-band. The friction coefficient from [18] is low compared to our experi-ment, probably due to the impact parameter selection in channeling conditions. To conclude, at velocities $v < 0.2$ a.u. excitation of sp-electrons is the dominant mechanism in electronic stopping of slow H ions in metals. Excitation of d-electrons only plays a role if the d-band extends up to $E_F$.

Fig. 2 displays the equivalent information for He ions. When $Q_{He}$ data are presented as a function of $r_{s,sp}$ (full symbols), the noble metals and Pt exhibit a strikingly different $r_s$-dependence than the DFT prediction. For the noble metals, the agreement between experiment and theory does not improve when using $r_{s,eff}$. As for H, stopping for C, Al, Zn and In is very well described by the FEG theory when using $r_{s,sp}$; for Pt, the use of $r_{s,eff}$ is appropriate. Also for He the TD-DFT data (asterisk) are low compared to our data, presumably again due to the impact parameter selection in channeling conditions as employed in [18]. In [18], it was observed that for Au the ratio $Q_{He}/Q_H$ exceeds predictions for $r_{s,sp}$ due to an enhanced participation of d-electrons in the interaction with He. In our experimental data such a behavior is observed for all noble metals.

As a next step, we want to investigate to which extent $Q_{He}$ is influenced by the d-band excitation threshold, $E_d$. To this end, we present the ratio $Q_{He}^{\exp}/Q_{He}^{FEG,sp}$ as a function of $E_d$ in Fig. 3a. This ratio clearly follows a trend: the relative importance of d-excitation is largest for $E_d = 0$ eV and decreases with increasing $E_d$, until it vanishes for $E_d \geq 8$ eV, thereby providing clear experimental evidence that for He ions excitation of d-electrons is already possible at ion velocities $v < 0.2$ a.u. (1 keV He) for $E_d \leq 8$ eV. The excitation efficiency is expected to depend also on the spatial distribution of the d-electrons. For instance, Cu and Au exhibit nearly identical excitation thresholds, but excitation of the

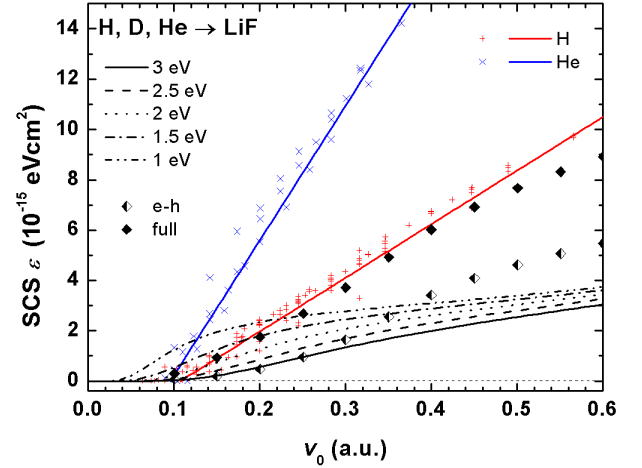

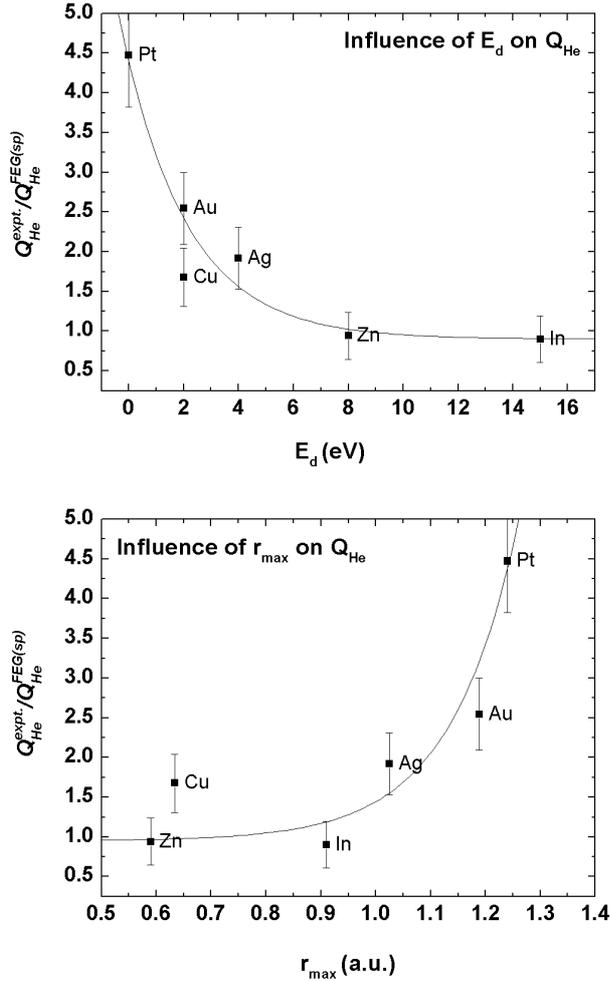

**FIG 3 (a)** *Friction coefficient ratio for He ions as a function of the d-band offset, $E_d$.* **(b)** *Friction coefficient ratio for He ions as a function of $r_{max}$, which corresponds to the maximum of the radial probability density of d-orbitals in the conduction band, obtained by atomic Hartree-Fock calculations [31]. Solid lines are to guide the eye.*

Au 5*d*-electrons is more likely due to their larger extension in space (see Fig. 3b).

In the second part of this investigation, we want to propose a possible solution to the puzzle why for large band gap insulators the observed threshold velocity $v_{th}$ is lower than predicted by theoretical models [19,20,32]. Since these models focus on electron hole pair excitations, we look for an additional process, which is already active at lower velocities. A possible candidate is energy loss by charge exchange cycles, a process that has already been observed in grazing scattering of H from LiF surfaces. After grazing surface collisions, scattered projectiles comprise a major fraction of H$^-$ [33,34], which was found to be correlated to energy dissipation by electron emission or exciton formation.

In the following we want to focus on LiF as a model system. In an ionic crystal, creation of H$^-$ is facilitated by the strong Madelung potential at the anion sites. When H$^0$ is close to F$^-$, the Madelung potential leads to a strong lowering of its affinity level and, consequently, to an increase of the probability for electron transfer from F$^-$ to H$^0$, $P_{bin}$. When H$^-$ approaches the next F$^-$ site, electron-hole pair or exciton formation is caused by Coulomb repulsion. In this mechanism, electron-hole pair excitation is not impeded by the large band gap. Such a charge exchange cycle contributes to the electronic stopping cross section per LiF molecule, namely by $\varepsilon_{CC} = (\Delta E \cdot P_{bin}/d_{FF}) \cdot V_{LiF}$. Here, $\Delta E$ denotes the energy loss per charge exchange cycle (~14eV), $d_{FF}$ is the distance between two F$^-$ sites (4.03 Å) and $V_{LiF}$ is the volume of a primitive unit cell. $P_{bin}$ can be evaluated using the Demkov model [35]:

$$P_{bin} = \frac{1}{2} \operatorname{sech}^2\left[\pi\gamma \frac{\Delta E_{FH}}{2v}\right] \quad (2)$$

**FIG 4** *Electronic stopping cross section of H in LiF as a function of ion velocity. Experimental data is represented by + (H) and x (He) symbols. Model calculations for the influence of negative ions are given by straight black lines for different values for $\Delta E_{FH}$. Semi-open diamonds indicate results of DFT calculations for electron-hole pair excitations [19]. Full symbols correspond to a combination of electron-hole pair excitation (DFT) and energy loss due to formation of transient negative ions ($\Delta E_{FH}$ = 2 eV).*

Here, $\Delta E_{FH}$ denotes the energy gap between the affinity level of H and the bound level of F$^-$; $\gamma$ is the decay constant of the exchange interaction [36]; in grazing scattering experiments, $\gamma = 1.7$ was used for the H-LiF system [37]. Additionally, one may consider the kinetic energy of the electron in H$^-$ when moving with velocity $v$. With this information, the following expression for the stopping cross section can be deduced:

$$\varepsilon_{cc} \sim (14 + 13.6 \cdot (v/v_0)^2) \cdot 4.06 \cdot 10^{-16} \cdot P_{bin} \text{ eVcm}^2 \quad (3)$$

In Fig. 4 experimental $\varepsilon$ data are compared to the results of model calculations that include contributions of electron-hole pair excitation (semi-open diamonds) and charge exchange (black lines). Electron-hole pair excitation is taken from recent DFT calculations [19]. The only open question concerns the magnitude for $\Delta E_{FH}$. Quantitative information on $\Delta E_{FH}$ is not available, and it may depend on the H-F distance. In Fig. 4, $\varepsilon_{CC}$ is shown for a number of fixed $\Delta E_{FH}$ values in the interval 1 eV ≤ $\Delta E_{FH}$ ≤ 3 eV. For $\Delta E_{FH}$ = 2 eV, the threshold behavior of H in LiF is well reproduced. When one adds the contributions due to electron-hole-pair excitations [19], $\varepsilon_{eh}$, excellent agreement with experimental data is obtained in the velocity range under consideration. This may be taken as a strong indication that the threshold behavior of $dE/dx$ is dominated by charge exchange cycles also in the bulk of an insulator and e-h pair excitation sets in at larger velocities. This model can also be applied to He projectiles. While for LiF, He exhibits the same threshold behavior as H, it features a significantly higher stopping power (see Fig. 4). This increased stopping efficiency cannot be explained by charge exchange cycles, since ΔE is limited to ~14 eV. Compared to H, however, He is expected to be more efficient in exciting e-h pairs, similarly as in metals.

Electronic stopping due to formation of transient negative ions can not only explain the observed threshold behavior in different ionic crystals, but also in oxides when they exhibit a sufficiently large Madelung potential. For instance, in $SiO_2$ the Madelung potential amounts to –15.3 eV at the $O^-$ sites [38], the band gap is $\Delta E_g \approx 8$ eV and the mean distance between O atoms is ~2.6 Å. The stopping thresholds for H and for He are well reproduced by reasonable choices for $\Delta E_{HO} \approx 1.25$ eV and $\Delta E_{HeO} \approx 0.5$ eV. In this way the puzzling fact can be explained that for $SiO_2$ the data point to a finite threshold only for H, not for He [21].

To summarize, electronic stopping of H and He has been analyzed at very low velocities (v < 0.2 a.u.). In metals, the *d*-electrons contribute with significantly different efficiency to electronic stopping of H and He ions. For H, the *d*-band participates in the stopping process only if it extends up to $E_F$; even Cu and Au can be adequately described by a FEG model which considers *sp*-electrons only. In contrast, He can excite deeper lying *d*-electrons, at least up to $E_d \approx 4$ eV.

In insulators, the experimentally observed thresholds in electronic stopping of H and He in large band gap insulators may be attributed to charge exchange cycles involving the formation of transient negative ions at anion sites. At higher velocities (*v* > 0.2 a.u.) electron-hole pair excitation will represent the prevailing mechanism of electronic energy loss.

### ACKNOWLEDGEMENT:

Inspiring discussions with Helmut Winter (HU Berlin) and Peter Zeppenfeld (JKU Linz) are gratefully acknowledged. We acknowledge support by the Austrian Science Fund (FWF): Project P22587. D Goebl was supported by a DOC fellowship of the Austrian Academy of Sciences.